\documentclass[lettersize,journal]{IEEEtran}
\usepackage[noadjust]{cite}
\usepackage{amsmath,amsfonts}
\usepackage{algorithmic}
\usepackage{array}
\usepackage{textcomp}
\usepackage{stfloats}
\usepackage{url}
\usepackage{verbatim}
\usepackage{graphicx}
\usepackage{float}
\usepackage{multirow}
\usepackage{caption}
\usepackage{subcaption}
\usepackage{makecell}
\usepackage{hhline}
\usepackage{xcolor}
\def\BibTeX{{\rm B\kern-.05em{\sc i\kern-.025em b}\kern-.08em
    T\kern-.1667em\lower.7ex\hbox{E}\kern-.125emX}}
\usepackage{balance}
\begin{document}
\title{Advancing Hybrid Quantum Neural Network for Alternative Current Optimal Power Flow}

\author{Ze~Hu,~\IEEEmembership{Student Member,~IEEE}, Ziqing~Zhu,~\IEEEmembership{Member,~IEEE}, Linghua~Zhu, Xiang Wei,~\IEEEmembership{Student Member,~IEEE}, Siqi~Bu,~\IEEEmembership{Senior Member,~IEEE}, Ka~Wing~Chan,~\IEEEmembership{Member,~IEEE}

\thanks{Manuscript created Feb, 2025.}}

\markboth{}{}

\maketitle

\begin{abstract}
Alternative Current Optimal Power Flow (AC-OPF) is essential for efficient power system planning and real-time operation but remains an NP-hard and non-convex optimization problem with significant computational challenges. This paper proposes a novel hybrid classical-quantum deep learning framework for AC-OPF problem, integrating parameterized quantum circuits (PQCs) for feature extraction with classical deep learning for data encoding and decoding. The proposed framework integrates two types of residual connection structures to mitigate the ``barren plateau" problem in quantum circuits, enhancing training stability and convergence. Furthermore, a physics-informed neural network (PINN) module is incorporated to guarantee tolerable constraint violation, improving the physical consistency and reliability of AC-OPF solutions. 
Experimental evaluations on multiple IEEE test systems demonstrate that the proposed approach achieves superior accuracy, generalization, and robustness to quantum noise while requiring minimal quantum resources.
\end{abstract}


\section{Introduction}

\subsection{Background}
\IEEEPARstart{A}{lternative} Current Optimal Power Flow (AC-OPF) is a critical optimization problem integral to the operations and strategic planning of various stakeholders—including system operators and electricity market participants—for the efficient and reliable planning and real-time control in power systems. This pervasive reliance highlights OPF's centrality in power systems; OPF influences economic activities exceeding ten billion dollars annually in the United States \cite{cain2012history, chiang2023dynamic}. 

One of primary objective of OPF is to minimize operational costs while satisfying physical constraints such as line capacities and bus voltages by determining the optimal power generation setpoints. However, the OPF problem is both NP-hard and non-convex, resulting in significant convergence challenges and protracted computational times for most solvers \cite{bienstock2019strong}. To address these challenges and ensure rapid computation with convergence guarantees, a substantial body of literature focuses on deriving linear or convex approximations of the AC-OPF problem, with the DC-OPF being one of the most prevalent methods \cite{kundur2007power}. 
Other methods like second-order cone approximation and relaxation \cite{halilbavsic2018convex} merit convexity and scalability but may suffer from approximation errors or yield solutions that are infeasible. Despite these advancements, AC-OPF still necessitates further improvements in computational efficiency and convergence speed.

\subsection{Related works}

In recent years, there has been a keen interest in employing machine learning methods, particularly neural networks (NNs), to estimate solutions of the AC-OPF \cite{duchesne2020recent, singh2021learning, zhou2022deepopf}. Compared to traditional approaches, NNs demonstrate strong performance in learning non-convex functions and have exhibited computational speedups more than 100 times \cite{nellikkath2022physics}. Moreover, multiple newly developed variants of NN are applied to address traditional challenges in OPF solutions like computational burden. For instance, by exploiting the graph nature of the power grid, \cite{owerko2020optimal} applied a graph neural network (GNN) to process node information locally in OPF problems, meriting better scalability. 
Despite their promising performance in OPF problems, NNs face several challenges in real-world implementation. First, NN performance in solving OPF depends heavily on large, high-quality training datasets, making them vulnerable to common machine learning issues such as data insufficiency or loss in available datasets. Second, training complexity grows rapidly with system size. As input and output dimensionality increases, NNs require more neurons and layers to model the system, which complicates the training process and demands substantial computational resources.

Quantum computing, leveraging principles of quantum mechanics like superposition, entanglement, and interference, offers a promising approach for solving complex problems and accelerating computations in both academia and industry \cite{morstyn2024opportunities}. Unlike classical computing, it operates on fundamentally different principles and is implemented through two main paradigms: adiabatic quantum computation (AQC) and gate-based quantum computation (GQC).
The former, AQC, solves problems by leveraging the adiabatic theorem of quantum mechanics \cite{kato1950adiabatic}, finding solutions by evolving a simple quantum system into a more complex one with small steps while remaining in the ground state throughout the whole process \cite{albash2018adiabatic}. On the other hand, GQC, considered the “standard” model of quantum computation \cite{deutsch1989quantum}, stores data in quantum bits (qubits) and operates on them using quantum gates within circuit models. A majority of quantum computation applications in power systems lie in the GQC paradigm \cite{eskandarpour2020quantum, saevarsson2022quantum, zhou2022noise, kaseb2024quantum}, especially since quantum devices first demonstrated computational advantages over classical devices \cite{cerezo2021variational}. For example, the Harrow–Hassidim–Lloyd algorithm (HHL) has been applied to solve the DC-OPF problem and achieved superior performance in small-scale systems \cite{eskandarpour2020quantum, saevarsson2022quantum}. Results highlighted the exponential speedup of HHL in solving a system of linear equations (SLE) over classical computers.
However, the performance of GQC methods is sensitive to their environment (noisy) and prone to quantum decoherence on the widely used noisy intermediate-scale quantum (NISQ) devices \cite{sagingalieva2023hybrid}. For their robustness to noise, parametrized quantum circuit (PQC), which has tunable parameters in the quantum gates and is optimized by a classical outer loop to solve a specific task, becomes a competitive candidate and applied in many practical scenarios \cite{cerezo2021variational}.
In \cite{zhou2022noise}, a high expressibility, low-depth (HELD) quantum circuit has been designed to solve the transient stability assessment in bulk power systems.

By incorporating PQC as layers within NN, quantum neural networks (QNNs) represent a new class of machine learning models that enhance classical methods by leveraging quantum effects such as superposition, entanglement, and interference \cite{abbas2021power}.  Notably, it has been demonstrated in \cite{beer2020training} that QNNs can generalize effectively from very small datasets and exhibit remarkable tolerance to noisy training data.
Furthermore, QNNs offer the ability to explore high-dimensional feature spaces by utilizing the tensor product structure of Hilbert spaces, thereby opening up possibilities for superior performance  \cite{lloyd2013quantum}. These advantages of QNN make it a strong candidate for non-convex power systems applications \cite{morstyn2022annealing}. For instance, \cite{kaseb2024quantum} adopted a hybrid QNN in power flow analysis, demonstrating a better generalization across systems.

\subsection{Research Gaps and Contributions}

Currently, state-of-the-art methods for solving the AC-OPF problem are still facing notable challenges. \textbf{First}, traditional mathematical optimization approaches struggle with scalability in large-scale power systems, failing to meet the speed and accuracy requirements of real-time operations. While promising, deep learning methods are constrained by the availability and quality of training data. The increasing complexity of modern power systems complicates the collection of high-resolution, real-time data, while uncertainties in renewable energy generation introduce noise into training samples. These factors result in unseen scenarios during deployment, demanding robust generalization from deep learning models. \textbf{Second}, quantum deep learning methods, although offering advanced capabilities in data representation and feature extraction, are hindered by the limitations of current quantum hardware. The scarcity of usable qubits restricts their application to large-scale AC-OPF problems. Additionally, the ``barren plateau" phenomenon, where gradients in high-dimensional parameter spaces diminish to near-zero during the training of PQC, further constrains the training and convergence of quantum circuits. More critically, qubits, due to their inherent physical characteristics, are highly susceptible to environmental noise, which can induce state-flipping errors, etc., leading to output inaccuracies. For AC-OPF, where accurate solutions are vital for the real-time operation of power systems, such errors can result in violations of critical physical constraints, e.g., line overloads or voltage limit breaches, so as to result in severe safety issues.

This paper proposes a novel hybrid classical-quantum deep learning model that leverages a small number of qubits to solve large-scale AC-OPF problems. This approach overcomes existing limitations in terms of accuracy, generalization, and quantum noise sensitivity. The main contributions of this paper are summarized as follows:

\begin{itemize}
	\item A novel integration of classical deep learning architectures with a customized PQC is proposed to solve the AC-OPF problem, enabling the hybrid classic-quantum methods to achieve high accuracy and robust generalization with only a small number of qubits. In the proposed method, classical deep learning components are specifically designed for data encoding and decoding, while PQCs are utilized for feature extraction.
	\item An innovative residual connection framework tailored for the hybrid classical-quantum architecture is proposed to address the gradient vanishing issue caused by the Barren Plateau during quantum circuit training. In particular, two types of residual connection structures are designed to enhance the training stability and convergence of QNN effectively.
	\item A physics-informed neural network (PINN) layer is introduced into the hybrid model. The proposed PINN module directly corrects the quantum computation outputs by embedding discrepancies in the Karush-Kuhn-Tucker (KKT) conditions representing violations of power system operational constraints into the loss function. This reduces errors caused by quantum noise and mitigates operational safety risks arising from AC-OPF solutions.
\end{itemize}

\subsection{Paper Organization}
The rest of the paper is constructed as follows: The formulation of the AC-OPF problem, along with its KKT conditions, is presented in Section II. The fundamentals of quantum computing and novel designs of QNNs, along with a composite quantum noise model, are introduced in Section III. The numerical results comparing the proposed approach with benchmark neural networks are shown in Section IV, highlighting the generation performance and robustness to the quantum noise. The paper is concluded in Section V.

\section{Problem formulation}

The AC-OPF problem is an optimization problem for system operators to schedule and dispatch resources in both day-ahead and real-time stages. According to the operation requirements, the objective of AC-OPF could be multiple, including generation cost minimization, line loss minimization, etc. The typical objective of minimizing the total cost of active power generation is used in the compact AC-OPF presented as (1).
\begin{subequations}
	\begin{align}    
		\min_{\mathbf{V}, \mathbf{G}} & \ \mathbf{c}^T \mathbf{G} \tag{1a} \label{1a} \\    
		\text{s.t.} & \ \mathbf{V}^T \mathbf{L}_l \mathbf{V} = {a}_l^T \mathbf{G} + {b}_l^T \mathbf{D}, \quad l =\{1,2,...,N_L\} : \rho_l \tag{1b} \label{1b} \\    
		& \ \mathbf{V}^T \mathbf{M}_m \mathbf{V} \leq {d}_m^T \mathbf{D} + f_m, \quad m = \{1,2,...,N_M\} : \mu_m \tag{1c} \label{1c} 
	\end{align} \label{eq:1} 
\end{subequations}

In \eqref{eq:1}, $\mathbf{c}^T$ is the combined linear cost terms for the active and reactive power generations, $\mathbf{G} = \begin{bmatrix} \mathbf{P}_g^T, \mathbf{Q}_g^T \end{bmatrix}^T$, $\mathbf{D} = \begin{bmatrix} \mathbf{P}_d^T, \mathbf{Q}_d^T \end{bmatrix}^T$. Voltage vector ${\mathbf{V}} = \begin{bmatrix} (\mathbf{V}^r)^T, (\mathbf{V}^i)^T \end{bmatrix}^T$ contains both the real and imagination part of voltage at all buses. $\mathbf{L}$ and $\mathbf{M}$ are combined vectors that stand for expressions of multiple constraints; please refer to \cite{nellikkath2022physics} for details. ${a}_l$ and ${b}_l$ are vectors mapping generation and demands to the corresponding buses. $d_m$ is the vector mapping active and reactive power of demand in $\mathbf{D}$ to corresponding constraints. $f_m$ stands for vectors of bounded values for voltage and line flows. $\rho_l$ and $\mu_m$ are dual variables for constraints \eqref{1b} and \eqref{1c}, respectively. $N_L$ is the number of constraints in \eqref{1b}, which is equal to $2N_b+1$.  $N_M$ is the number of constraints in \eqref{1c}, which is equal to 4 times number of generators $4N_g+2N_b+N_l$. The objective (1a) is to minimize the generation cost, subject to compact form constraints of (1b) and (1c). (1b) denotes all equality constraints, including $2N_b$ constraints for bus power injection and $1$ constraint for angle reference.  (1c) stands for all inequality constraints, including $4N_g$ constraints for optimal power generation, $2N_b$ constraints for bus voltage, and $N_l$ constraints for line current flow. The Lagrangian function for the AC-OPF can be formulated as (2).
\begin{equation} \begin{array}{lrl}
		\mathcal{L}(\mathbf{x}, \rho, \mu, \mathbf{D}) &= \mathbf{c}^T \mathbf{G} + \sum_{l=1}^{L} \rho_l \left(\mathbf{V}^T \mathbf{L}_l \mathbf{V} - \mathbf{a}_l^T \mathbf{G} - \mathbf{b}_l^T \mathbf{D}\right) \\
		&\quad + \sum_{m=1}^{M} \mu_m \left(\mathbf{V}^T \mathbf{M}_m \mathbf{V} - \mathbf{d}_m^T \mathbf{D} - f_m\right) \tag{2}
\end{array}\end{equation}

Given the optimization problem and its Lagrangian function, the KKT conditions are presented in (\ref{1b})-(\ref{1c}) and \eqref{eq:KKT1}-\eqref{eq:KKT4}, where the stationarity condition is given in \eqref{eq:KKT1} and \eqref{eq:KKT2}, the complementary slackness condition and dual feasibility are given by \eqref{eq:KKT3} and \eqref{eq:KKT4}, respectively. 

\begin{subequations}
	\begin{align}
		\mathbf{c} &= \sum_{l=1}^{L} \rho_l \mathbf{a}_l \tag{3} \label{eq:KKT1}\\            
		\left(\sum_{l=1}^{L} \rho_l \mathbf{L}_l + \sum_{m=1}^{M} \mu_m \mathbf{M}_m\right) &= 0 \tag{4} \label{eq:KKT2}\\                
		\mu_m \left(\mathbf{V}^T \mathbf{M}_m \mathbf{V} - \mathbf{d}_m^T \mathbf{D} - f_m\right) &= 0, \quad m = \{1,2,...,M\}  \tag{5} \label{eq:KKT3}\\
		\mu_m &\geq 0, \quad m = \{1,2,...,M\} \tag{6} \label{eq:KKT4}\\                                    
		(\ref{1b})-(\ref{1c})\notag
	\end{align}
\end{subequations}

In practical scenarios, grid operators often need to address stochastic AC-OPF multiple times within a specified time window. However, the inherent NP-hard and nonconvex nature of the AC-OPF poses significant computational challenges. 

\section{Design of Hybrid Quantum Neural Network for AC-OPF}

This section introduces the fundamentals of quantum deep learning and proposes a novel integration of classical deep learning with PQC to solve the AC-OPF problem. Furthermore, an innovative residual framework tailored for hybrid classical-quantum architecture and a PINN layer designed to integrate constraints to the NN learning are developed. Therefore, two novel methods, physics-informed QNN with sequential residual connection (QNN-SRC) and physics-informed QNN with nested residual connection (QNN-NRC), are proposed. Finally, a comprehensive quantum noise model is developed to simulate the impact of quantum noise.

\subsection{Fundamentals of Quantum Neural Networks}

\begin{figure}[!ht]
	\centering
	\includegraphics[width=4.6cm]{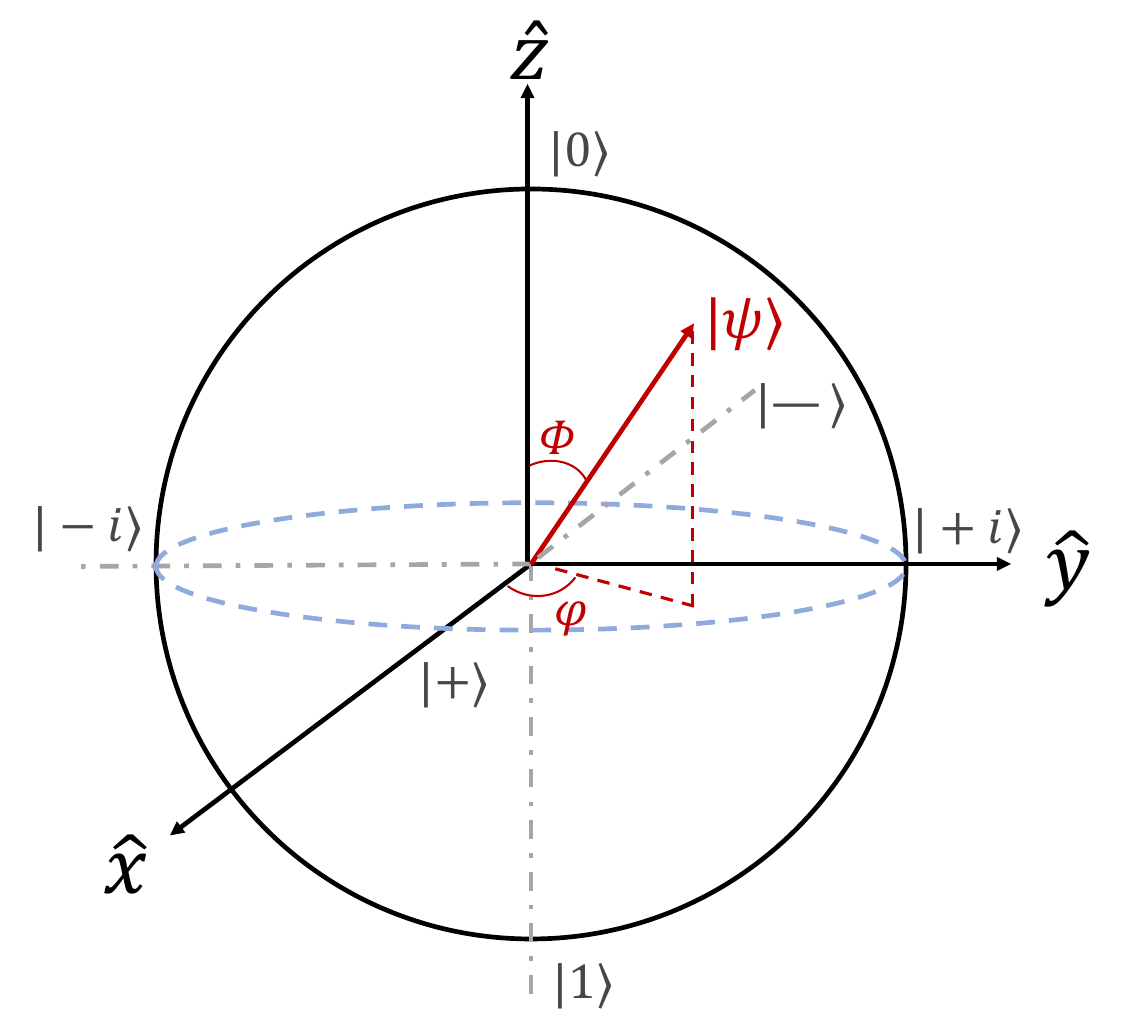} 
	\caption{Illustration of a qubit}
	\label{fig:qubit}
\end{figure}

A qubit, fundamentally defined by complex probability amplitudes, can be represented on the Bloch sphere using two angular coordinates. The schematic representation of a qubit is shown in Fig.~\ref{fig:qubit}. Unlike a classical bit that resides definitively in either state 0 or 1, a qubit inhabits a quantum superposition of both states until it is measured. This means the qubit's state is a linear combination of the basis states $|0\rangle$ and $|1\rangle$, introducing inherent probabilistic behavior. 

\begin{figure*}[!ht]
	\centering
	\includegraphics[width=12 cm]{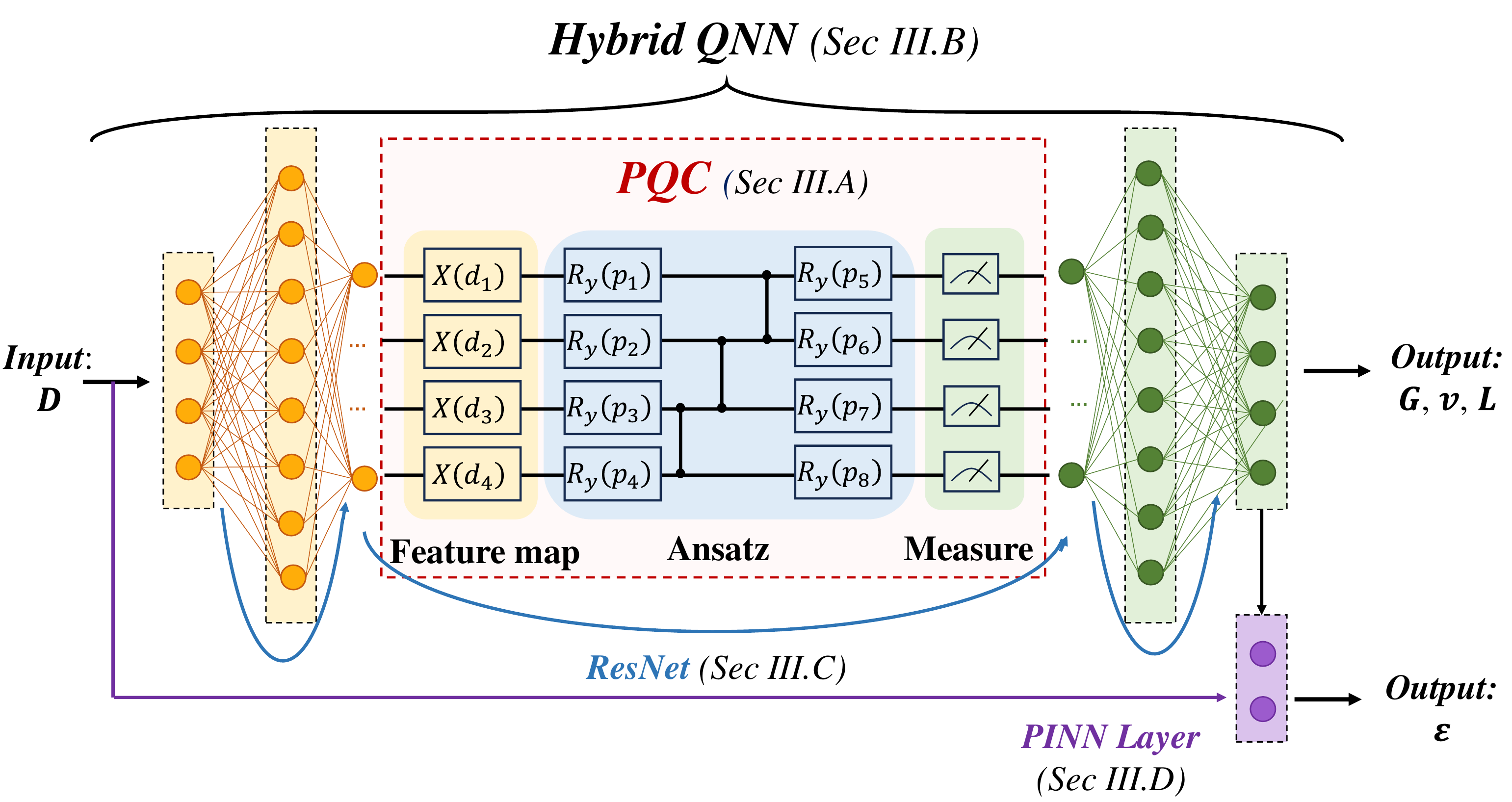} 
	
	\caption{Structure of a Hybrid QNN using residual connection and physics-informed layer}
	\label{fig:QNN}
\end{figure*}

QNNs are built as PQC within the continuous-variable (CV) framework, where information is encoded in continuous parameters like the amplitudes of electromagnetic fields. As depicted in Fig.\ref{fig:QNN}, a QNN constitutes feature mapping, an ansatz, and a measurement. Much like classical NNs, QNNs can be trained using optimization algorithms to learn the mapping between input features $\vec{x}$ and output labels $\vec{y}$. 

Within the PQC, two primary types of quantum gates are employed: 1) the rotation operator gate $R$ and 2) the controlled NOT (CNOT) gate. The rotation operator gates, $X(\theta)$ and $R_y(\theta)$, represent rotation matrices along two Cartesian axes, where $\theta$ denotes the rotation angle. These matrices are defined in equations \eqref{eq:x} and \eqref{eq:y}, illustrating rotations along the corresponding axes. The angle $\theta$ specifies the extent of rotation applied along each axis.
\begin{equation} 
	X(\theta) = 
	\begin{bmatrix}
		\cos(\theta/2) & -i \sin(\theta/2) \\
		-i \sin(\theta/2) & \cos(\theta/2)
	\end{bmatrix}
	\tag{7}
	\label{eq:x}
\end{equation} 
\begin{equation} 
	R_y(\theta) = 
	\begin{bmatrix}
		\cos(\theta/2) & -\sin(\theta/2) \\
		\sin(\theta/2) & \cos(\theta/2)
	\end{bmatrix}
	\tag{8}
	\label{eq:y}
\end{equation} 

The connection lines in the ansatz correspond to CNOT gates, which act on two qubits. The CNOT gate performs a NOT operation on the second qubit only when the first qubit is in the state $\lvert 1 \rangle$.
The training process begins with the feature mapping, where the input data $\vec{x}$ is encoded into the quantum states of multiple qubits. After encoding, QNN applies sequences of quantum gates with tunable parameters to form an ansatz—a trial solution that the network adjusts to minimize the loss function. Then, measurements are taken from the qubits. These measurements are then processed classically to generate predictions or labels, which are evaluated against the actual outputs using a loss function.
The representation of QNNs can be expressed as follows.
\begin{subequations} 
	\begin{align}
		&|\psi^{\text{in}}\rangle = U(\vec{x})|0 \dots 0\rangle, \tag{9a}\\ &|\psi^{\text{out}}\rangle = \omega(\vec{w}^r)|\psi^{\text{in}}\rangle\tag{9b}
	\end{align}
\end{subequations}

The input state $|\psi^{\text{in}}\rangle$ is generated by transforming $\vec{x} \in {\vec{p}_i, \vec{q}_i : i = 1, 2, \dots, n}$ into valid quantum states using the feature map $U(\cdot)$, which is applied to the vacuum state $|0 \dots 0\rangle$. The ansatz $\omega(\cdot)$ is then applied to $|\psi^{\text{in}}\rangle$. Notably, the feature map does not involve any parameters that require optimization. In contrast, the vector of adjustable parameters for the ansatz, denoted by $\vec{w}^r$, is fine-tuned during the training process using a dataset comprising $N$ training pairs ${\vec{x}, \vec{y}}$. The resulting output state $|\psi^{\text{out}}\rangle$ cannot be directly observed; instead, it must be inferred through measurement, yielding $\hat{y} \in {(\vec{v}_i, \vec{\delta}_i) : i = 1, 2, \dots, n}$.
Specifically, the optimization of the QNN involves adjusting the parameters within the PQCs to minimize the loss function, which is typically achieved using gradient-based optimization methods.

\subsection{Hybrid Quantum Neural Networks}
Hybrid QNN is a neural network that integrates a PQC as an intermediate layer within a traditional NN, as depicted in Fig. \ref{fig:QNN}. Data flow initiates from the input layer and the first hidden layer of the NN, traverses through the feature map and ansatz within the PQC, and, following measurement, proceeds to subsequent classical hidden layers and output layer. Afterward the final layer, $\mathbf{G}, \mathbf{V}, \mathbf{L}$ are output as the solution results. Notably, hybrid QNNs emulates a encoder-decoder structure, where the initial classical component encodes the input $x$ into a quantum latent space on PQC. Subsequently, the second classical component deciphers the measurement outcomes back into a higher-dimensional space to produce the output.

To train the Hybrid QNN with back-propagation to improve the solution accuracy for the AC-OPF, the loss function \( L_{\text{nn}} \) is represented by \eqref{eq:lag}, containing Mean Absolute Error (MAE) of active and reactive power generation \( G \), bus voltage \( V \), and Lagrangian multipliers \( L \). \( \Lambda \) are predefined parameters to weight terms. $N_t$ is the number of training instance. 
\begin{equation}
	L_{\text{nn}} = \frac{1}{N_t} \sum_{i=1}^{N_t} (\Lambda_P \frac{|\hat{G} - G|}{\text{MAE}_g} + \Lambda_V \frac{|\hat{V} - V|}{\text{MAE}_v} + \Lambda_L \frac{|\hat{L}_m - L_m|}{\text{MAE}_l} ) \tag{10} \label{eq:lag}
\end{equation}

\subsection{Residual Learning (ResNet)}

Residual learning, which leverages shortcut connections between layers in NNs, was initially introduced for image recognition tasks \cite{he2016deep}. By mitigating vanishing and exploding gradient problems, residual learning substantially enhances the performance of NNs across diverse tasks; moreover, it may also deal with the barren plateaus problem caused by the quantum noise \cite{qi2023barren}. Drawing inspiration from \cite{he2016deep,li2020encoder}, we propose two variants of residual-connected hybrid QNNs by applying two different kinds of shortcut connections, which are elaborated as follows. 


\subsubsection{Sequential Residual Connection}
\begin{figure*}[!ht]
	\centering
	\begin{subfigure}[b]{0.4\textwidth}
		\includegraphics[width=6.6cm]{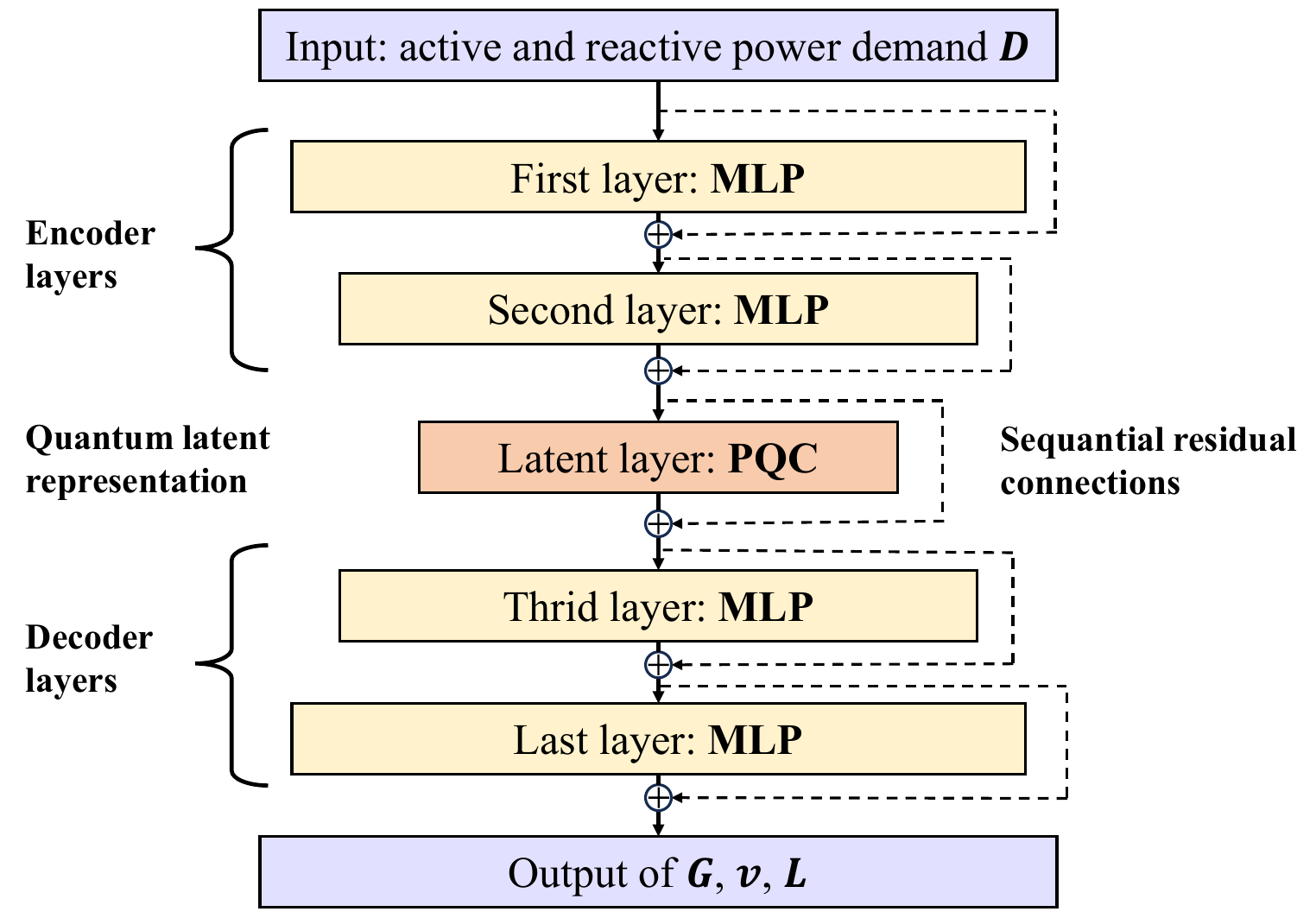} 
		\caption{Hybrid QNN with sequential residual connection}
		\label{fig: s_residual}
	\end{subfigure}
	\hspace{1cm}
	\begin{subfigure}[b]{0.4\textwidth}
		\includegraphics[width=6.6cm]{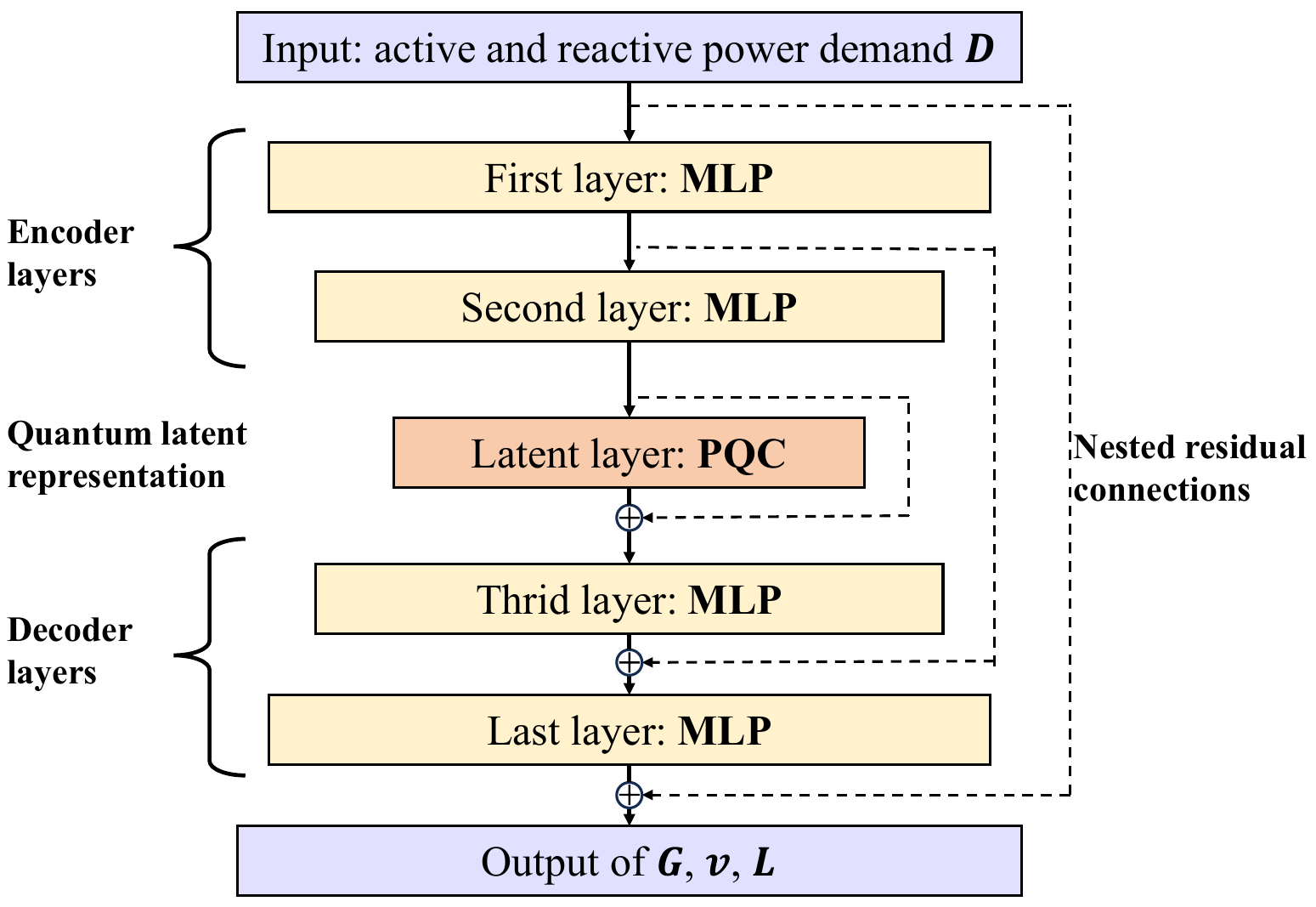} 
		\caption{Hybrid QNN with nested residual connection}
		\label{fig: n_residual}
	\end{subfigure}
	\caption{Structure of Hybrid QNN with two types of residual connections}
\end{figure*}

Sequential connections enable residual shortcuts from the input of the first layer to the output of the second layer before the activation function \cite{he2016deep}. It assumes that the whole NN gradually processes different features of the input. The detailed implementation is shown in Fig.~\ref{fig: s_residual} and represented by \eqref{s_r}. 

\begin{equation}
	y_{l} = x_l + f_l(x_l, W_l)
	\label{s_r}
	\tag{11}
\end{equation}

In \eqref{s_r}, $x_l$ and $y_l$ are the input and output for layer $l$, respectively. $W_l$ denotes the parameters for the layer $l$. In addition, the layer output $y_l$ is input to an activation function, e.g., rectified linear unit (ReLU), for non-linear transformation and then be the input of the next layer, which is shown in \eqref{s_r_2}.
\begin{equation}
	x_{l+1} = Relu(y_l)
	\tag{12}
	\label{s_r_2}
\end{equation}

\subsubsection{Nested Residual Connection}
The "nested" residual connections indicate the residual connection from the encoder to its corresponding "mirror" layer in the decoder. This kind of connection implies an assumption that these mirror layers encode or decode the same level of features in the NN. The implementation is presented in Fig.~\ref{fig: n_residual}. Assuming the PQC as the latent layer \(Q\), the encoder layer \(1\), and its mirror decoder layer \( L \). With the residual identity connection, it can be presented as
\begin{equation}
	y_L = x_l + f_L(x_L, W_L)
	\tag{13} \label{eq:yl1}
\end{equation}
where \( x_l \) and \( y_l \) denote the input and output of the \( l \) layer, \( x_L \) and \( y_L \) are the input and output of the \( L \) layer. The \( W_l \) and \( W_L \) indicate parameters for the corresponding \( l \) and \( L \) layers. \( f_L() \) is the layer function including activation. As the decoder layer is behind the encoder and also the latent layer, equation \eqref{eq:yl1} can be reformulated by \eqref{eq:yl2}, where \( q_L(f_L(x_l, W_l), W_q) \) stands for the function of encoder input \( x_l \) for decoder output \( x_L \), indicating multi-layer transformation.
\begin{equation}
	y_L = x_l + f_L(q_L(f_L(x_l, W_l), W_q), W_L)
	\tag{14} \label{eq:yl2}
\end{equation}

From automatic differentiation \cite{li2020encoder}, the general derivative of the loss function for \( x_l \) can be used to compute the gradients for the parameters of the previous layer as \eqref{eq:diflag}, where the activation function and batch normalization are omitted for simplicity but are added to improve representation ability and avoid the "vanishing gradient" problem in implementation.
\begin{equation}
	\frac{\partial L_{nn}}{\partial x_l} = \frac{\partial L_{nn}}{\partial y_L} \cdot \left( 1 + \frac{\partial}{\partial x_l} \left( f_L \left( g_L \left( f_l(x_l, W_l) \right) , W_L \right) \right) \right)
	\tag{15} \label{eq:diflag}
\end{equation}

The constant \( 1 \) in the loss function above directly propagates the data from the decoder to the corresponding encoder layer. Also, the second term after the constant 1 is not always \(-1\), which may cancel out the gradient. Therefore, gradient vanishing and subsequent degradation of performance can be avoided during backward propagation.

\subsection{Physics-informed Neural Network}

\begin{figure}[!ht]
	\centering
	\vspace{-4mm}
	\includegraphics[width=6.6cm]{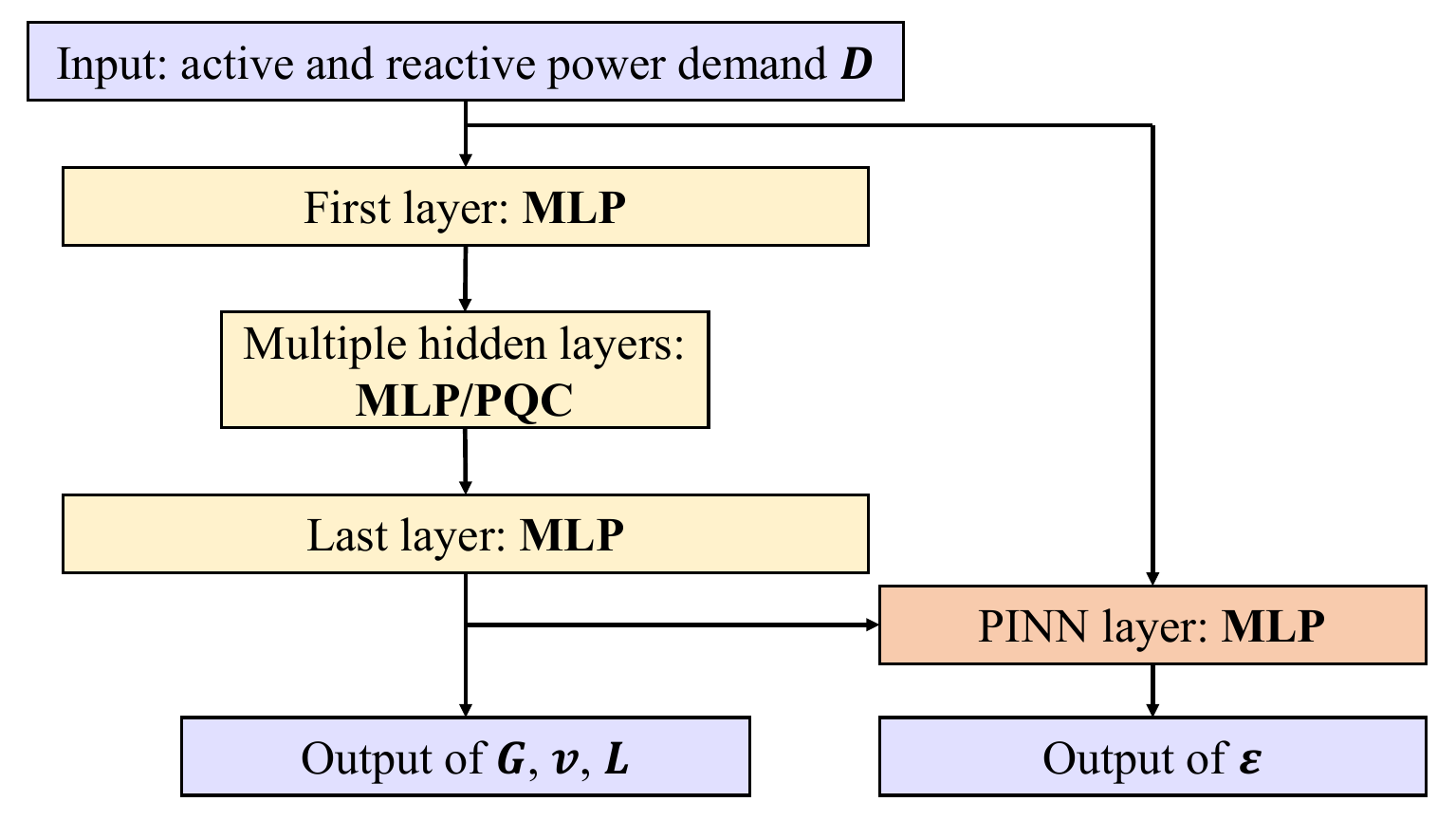} 
	\caption{Detailed illustration of a Hybrid Quantum PINN}
	\vspace{-4mm}
	\label{fig:pinn}
\end{figure}

In this subsection, a PINN layer is designed to integrate constraints to the NN learning, aiming to mitigate operational safety risks in power systems arising from AC-OPF solutions.
As illustrated in Fig.\ref{fig:pinn}, PINNs integrate an additional layer after the NN output to estimate the differential terms of the output variables, representing the "physics loss", minimizing which can lead to solutions against physics laws' violation.

In the context of the AC-OPF problem, the optimal solution must satisfy the KKT conditions specified in (\ref{1b})-(\ref{1c}) and \eqref{eq:KKT1}-\eqref{eq:KKT4}, the discrepancies in the KKT conditions—as shown in \eqref{eq:eps1} to \eqref{eq:eps4}—represent the physical loss in the AC-OPF. 
\begin{align}&\epsilon_{\text{stat}} = \left| c - \sum \hat{\rho_l} a_l \right| + \left| \left( \sum \hat{\rho_l} L_l + \sum \hat{\mu_m} M_m \right) \right|     \tag{16-a} \label{eq:eps1}\\
	&\epsilon_{\text{comp}} = \sum \left| \hat{\mu_m} \left( \hat{V}^T M_m \hat{V} - d_m^T D - f_m \right) \right|  \tag{16-b} \label{eq:eps2}\\
	&\epsilon_{\text{dual}} = \pi(\hat{\mu_m})  \tag{16-c} \label{eq:eps3}\\
	&\epsilon_{\text{prim}} = \sum \left| \hat{V}^T L_l \hat{V} - a_l^T G - b_l^T D \right| \tag{16-d}    \label{eq:eps4}
\end{align}

By combining the mean absolute errors (MAEs) for $G$, $v$, and $L$,as well the discrepancy in KKT conditions, the MAE in \eqref{eq:MAE} is used to be the loss function and update the network, where $N_c$ represents the number of collection instant.

\begin{equation}\begin{array}{lrl}
		MAE = \frac{1}{N_t} \sum_{i=1}^{N_t} ( \Lambda_P \underbrace{|\hat{\mathbf{G}} - \mathbf{G}|}_{MAE_g} + \Lambda_V \underbrace{|\hat{\mathbf{V}} - \mathbf{V}|}_{MAE_v} + \\      \Lambda_L \underbrace{|\hat{\mathbf{L}}_m - \mathbf{L}_m|}_{MAE_l})       + \frac{\Lambda_\epsilon}{N_t + N_c} \sum_{i=1}^{N_t + N_c} \underbrace{\epsilon_{\text{stat}} + \epsilon_{\text{comp}} + \epsilon_{\text{dual}} + \epsilon_{\text{prim}}}_{MAE_\epsilon}  \tag{17} \label{eq:MAE}
\end{array}\end{equation}

Furthermore, collocation points are incorporated into the training set to train the PINN layer by leveraging the KKT conditions to assess the accuracy of the neural network's predictions. This technique has been validated to enhance the training efficiency of PINNs \cite{nellikkath2022physics}. Specifically, collocation points are a set of random input values from the input domain akin to the NN training data points. However, only the error terms specified in equations (11) are utilized to measure prediction accuracy and train the NN; the optimal generation dispatch values, voltage set-points, and dual variables are not computed for the collocation points prior to training. 

\subsection{Composite Noise Model}
To realistically simulate noise in quantum systems, a composite noise model is constructed by incorporating four fundamental error channels: depolarizing noise, amplitude damping, readout errors, and coherent over-rotations. 

\subsubsection{Depolarizing Noise}
Depolarizing noise represents random Pauli errors affecting quantum states. For a single-qubit channel, the noise transformation is given by \eqref{eq:depo}, where $\rho$ is the density matrix representing the quantum state of a system, $e_s$ and $e_t$ denote the depolarizing probabilities for single-qubit gates (e.g., \texttt{u3}) and two-qubit gates (e.g., \texttt{cx}), respectively. $X, Y, Z$ are the Pauli matrices, which represent quantum operations affecting a single qubit. This noise effectively randomizes the quantum state, mimicking decoherence effects encountered in superconducting qubit systems.

\begin{equation}
	\rho \rightarrow (1-e_s)\,\rho + \frac{e_t}{3}\left(X\rho X + Y\rho Y + Z\rho Z\right)\tag{18}
	\label{eq:depo}
\end{equation}

\subsubsection{Amplitude Damping}
Amplitude damping models energy relaxation, where an excited state $\lvert 1 \rangle$ decays to the ground state $\lvert 0 \rangle$ with probability $e_d$. This process reflects interactions between the qubit and its environment, such as spontaneous emission. Mathematically, the Kraus operators $K_0, K_1$ describing the probability of state decay for amplitude damping are in \eqref{eq:ampl}. This noise particularly affects gates like \texttt{u2}, where state superpositions are sensitive to relaxation processes.

\begin{equation}  
	K_0 = \begin{pmatrix} 1 & 0 \\ 0 & \sqrt{1-e_d} \end{pmatrix}, \quad  
	K_1 = \begin{pmatrix} 0 & \sqrt{e_d} \\ 0 & 0 \end{pmatrix} \tag{19}
	\label{eq:ampl}
\end{equation}

\subsubsection{Readout Noise}
Readout noise accounts for measurement errors, where the recorded outcome differs from the actual qubit state. This is modeled using a misclassification probability $e_m$, leading to a transition matrix as \eqref{eq:read}. This error reflects imperfections in detector fidelity and signal processing, occurring uniformly across all measured qubits.
\begin{equation}
	M = \begin{pmatrix} 1-e_m & e_m \\ e_m & 1-e_m \end{pmatrix}
	\tag{20}
	\label{eq:read}
\end{equation}

\subsubsection{Coherent Errors}
Coherent errors arise from systematic imperfections, such as gate over- or under-rotations. A small over-rotation about the $X$-axis by an angle $e_c$ radians is represented by \eqref{eq:cohe}. Unlike stochastic noises above, coherent errors accumulate over repeated operations, leading to long-term deviations in quantum circuit behavior. These errors typically affect parameterized gates like \texttt{u3}, resulting from imperfect hardware calibration.
\begin{equation}
	U_{\text{error}} = \cos\left(\frac{e_c}{2}\right)I - i\sin\left(\frac{e_c}{2}\right)X
	\tag{21}
	\label{eq:cohe}
\end{equation}

\section{Numerical results}

\subsection{Simulation setup}

The proposed method is evaluated against a standard NN and PINN across several standard IEEE systems. The comparison methods include a conventional NN, PINN, Hybrid QNN, QNN-SRC, and QNN-NRC. All QNN-based methods mentioned above are realized based on a 4-qubit PQC. We used Pytorch \cite{paszke2019pytorch} with Python for NN training and Qiskit \cite{qiskit2024} for the implementation of PQC. To achieve a better performance, the input of NN, including active and reactive power demand, is normalized to have a mean of 0 and a standard deviation of 1. The number of training epochs is fixed to be 1,000, and the data set is split into 50 batches during training.

Ten thousand sets of random active and reactive power input values were generated using Latin hypercube sampling \cite{mckay2000comparison}. Among these, $50\%$ were allocated to the collocation data set, for which OPF set-points were not required, $20\%$ was used for training, while $30\%$ served as the testing set. AC-OPF in MATPOWER was utilized to determine the optimal active and reactive power generation values and voltage setpoints for the input data points in the training and test sets.
The network training is conducted on a PC with Intel(R) Xeon(R) Gold 6230R CPU@ 2.10GHz 2.10 GHz, 256 GB RAM, and NVIDIA RTX A4900. 


\subsection{Network Convergence in Training Set}

\begin{figure}[!ht]
	\centering
	\vspace{-4mm}
	\includegraphics[width=6.6cm]{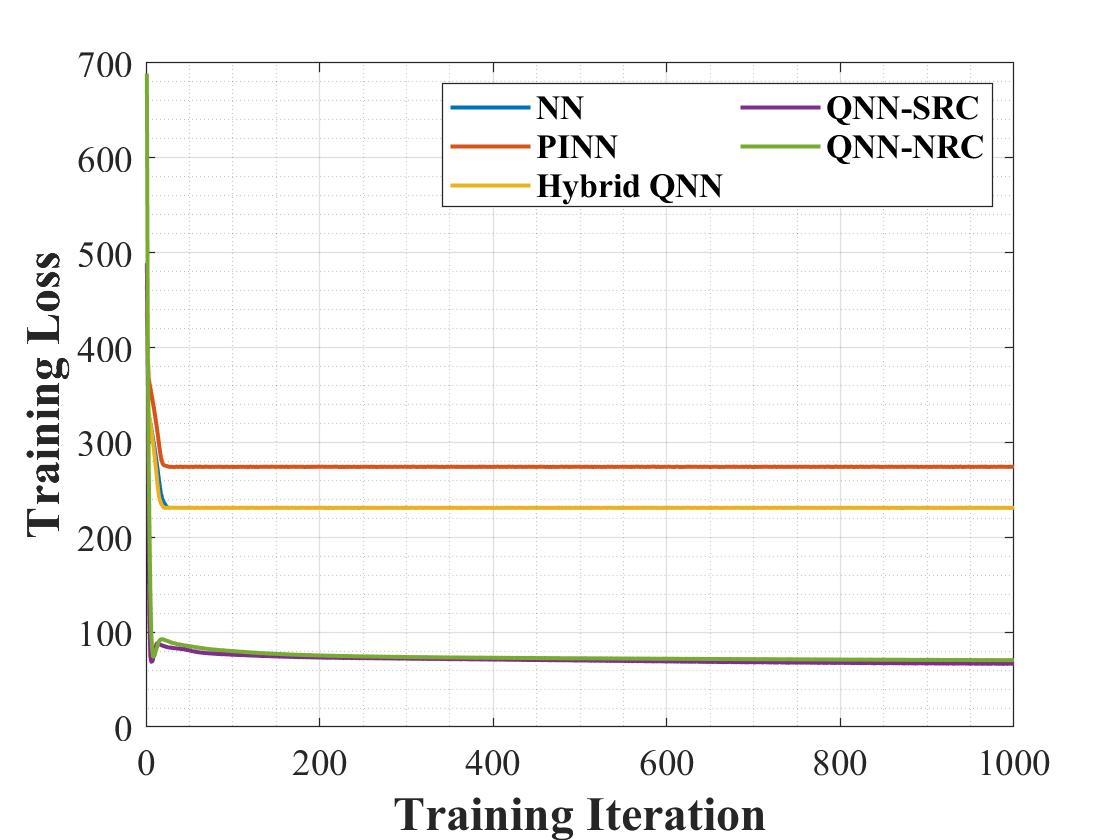}
	\caption{Training loss curves for comparable methods on the IEEE 39-bus system}
	\vspace{-2mm}
	\label{fig:y equals x}
	\label{converge}
\end{figure}

Fig.~\ref{converge} presents the training loss curves of various neural network-based methods for solving the OPF problem on the IEEE 39-bus system. Initially, all methods exhibit high loss values, which rapidly decrease within the first few iterations. Following this phase, each method converges along a distinct trajectory.
Among the methods, QNN-SRC and QNN-NRC achieve the lowest final loss values, demonstrating superior performance. Notably, both exhibit an early abrupt drop, suggesting efficient learning dynamics, with QNN-SRC stabilizing at a slightly lower loss than QNN-NRC. In contrast, NN and QNN converge to similar, moderately higher loss levels, indicating their limitations compared to residual-based approaches. 
The PINN consistently yields the highest loss, likely due to the additional KKT loss term, which enforces OPF constraints but complicates optimization. These results highlight the effectiveness of proposed approaches over existing methods (NN, PINN, Hybrid QNN) in achieving lower loss and better convergence in OPF applications.

\begin{figure}[!ht]
	\centering
	\vspace{-2mm}
	\includegraphics[width=7.6cm]{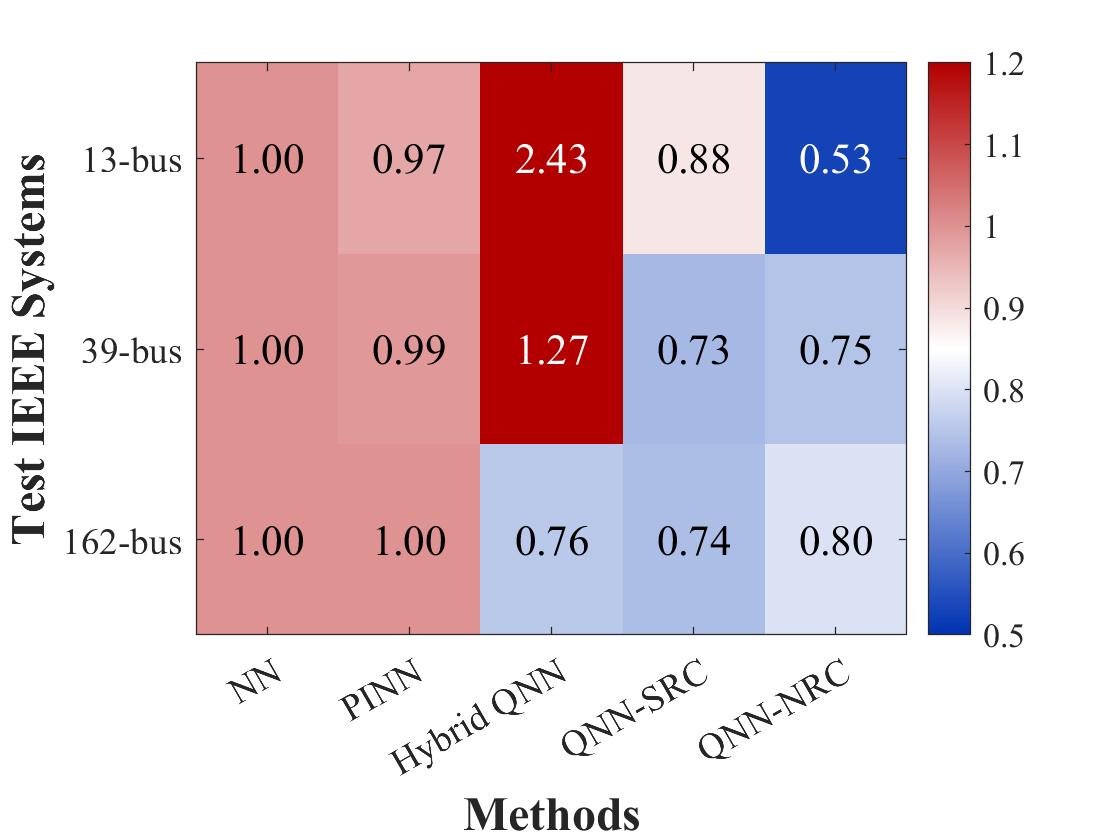} 
	\caption{Total MAEs comparison across test systems and methods}
	\vspace{-4mm}
	\label{heatmap}
\end{figure}

\subsection{Evaluation of Methods Performance Across Test Systems}

This subsection evaluates the effectiveness of the proposed methods across different test systems, specifically the IEEE 14-bus, 39-bus, and 162-bus systems. 
The heat map in Fig.~\ref{heatmap} presents the normalized total MAE for each method across the test systems, computed as the sum of all individual MAEs normalized by total MAEs of NN. Lower total MAEs, represented by blue regions, indicate superior model performance, whereas higher losses, shown in red, reflect poorer convergence and accuracy. 
Several key observations can be made from this visualization. First, QNN-SRC and QNN-NRC consistently achieve the lowest total loss values, particularly in the IEEE 39-bus and 162-bus systems, suggesting their superior strong generalization ability over comparable methods. NN and PINN exhibit relatively similar loss values across all systems, indicating that incorporating a physics-informed layer in PINN provides only marginal improvements over a standard neural network. Hybrid QNN demonstrates the highest total loss in the IEEE 14-bus system, indicating that a Hybrid QNN without additional enhancements does not necessarily offer advantages over conventional methods. The sharp contrast between Hybrid QNN’s performance in the 14-bus system, where it has the highest loss, and in the 162-bus system, where its loss is lower, suggests that its effectiveness is system-dependent.

Table~\ref{sample_table} provides a detailed breakdown of MAEs for active and reactive power set-points, as well as bus voltage magnitude and angle, across all models and test systems, offering further insights into individual metric performance.
In the IEEE 14-bus System, QNN-SRC achieves the lowest MAEs for real power set-points and voltage magnitude, while QNN-NRC achieves the lowest MAEs for reactive power set-points and voltage angles. Compared to conventional methods, the proposed models reduce MAEs by approximately $5\%–70\%$, demonstrating substantial improvement.
In the IEEE 39-bus System, QNN-SRC outperforms all other methods in estimating active and reactive power set-points and voltage angles, achieving MAE reductions of $30\%$, $17\%$, and $35\%$, respectively. The best voltage magnitude estimation is obtained using QNN-NRC, which reduces the MAE by $20\%$.
In the IEEE 162-bus System, QNN-SRC achieves the highest accuracy in estimating active power set-points, while Hybrid QNN provides the most precise estimations for reactive power set-points and voltage angles. NN achieves the lowest MAE in voltage magnitude estimation, highlighting that larger systems may exhibit different trends in performance. Although the proposed models do not necessarily achieve the lowest MAEs for each individual metric, their total MAE, computed as the sum of all individual MAEs, remains the lowest, particularly due to significant reductions in active power MAE.

Overall, the results indicate that QNN-SRC and QNN-NRC consistently outperform conventional NN-based approaches across different test systems, particularly in achieving lower total loss and superior MAEs in key metrics.

\begin{table}[!t]
	\renewcommand{\arraystretch}{1.3} 
	\caption{MAEs comparison across different test systems}
	\label{sample_table}
	\setlength{\tabcolsep}{3.6pt}
	\centering
	\begin{tabular}{cccccc}
		
		\hline
		\textbf{IEEE Systems} &	\textbf{Models} & \textbf{MAE($P$)} & \textbf{MAE($Q$)}  & \textbf{MAE($V$)}& \textbf{MAE($\theta$)} \\ \hline
		\multirow{5}{*}{\makecell{\textbf{14-bus}}}
		&NN&0.003845	&0.009721	&0.000967	&0.002928
		\\
		&PINN&0.003383	&0.009706	&0.000955	&0.002927\\
		&Hybrid QNN&0.024254	&0.012136	&0.001424	&0.004699\\
		&QNN-SRC&\textbf{0.002991}	&0.008540	&\textbf{0.000936}	&0.002957\\
		&QNN-NRC & 0.003136	&\textbf{0.002654}	&0.000987	&\textbf{0.002466}  \\ \hline
		\multirow{5}{*}{\textbf{39-bus}} 
		&NN &0.233563	&0.074443	&0.005198	&0.016926\\
		&PINN &0.231455	&0.073977	&0.005289	&0.016411\\
		&Hybrid QNN &0.320735	&0.076275	&0.00452	&0.017960\\
		&QNN-SRC &\textbf{0.163985}	&\textbf{0.061558}	&0.004040	&\textbf{0.011216}\\
		&QNN-NRC & 0.167899	&0.062898	&\textbf{0.003256}	&0.012744  \\ \hline
		\multirow{5}{*}{\textbf{162-bus}}
		&NN   & 34.02652	&15.62813	&\textbf{0.030486}	&1.068521	 \\ 
		&PINN   & 34.02775	&15.62901	&0.030526	&1.068825 \\ 
		&Hybrid QNN  & 23.44622	&\textbf{14.40025}	&0.046332	&\textbf{0.509803}\\ 
		&QNN-SRC     & \textbf{21.88830}	&14.49269	&0.142383	&0.814010 \\ 
		&QNN-NRC & 24.89359	&14.60358	&0.182287	&0.835687  \\ \hline
	\end{tabular}
\end{table}

\subsection{Sensitivity Analysis on Quantum Noise}
\begin{figure}[!ht]
	\centering
	\vspace{-4mm}
	\includegraphics[width=6.6cm]{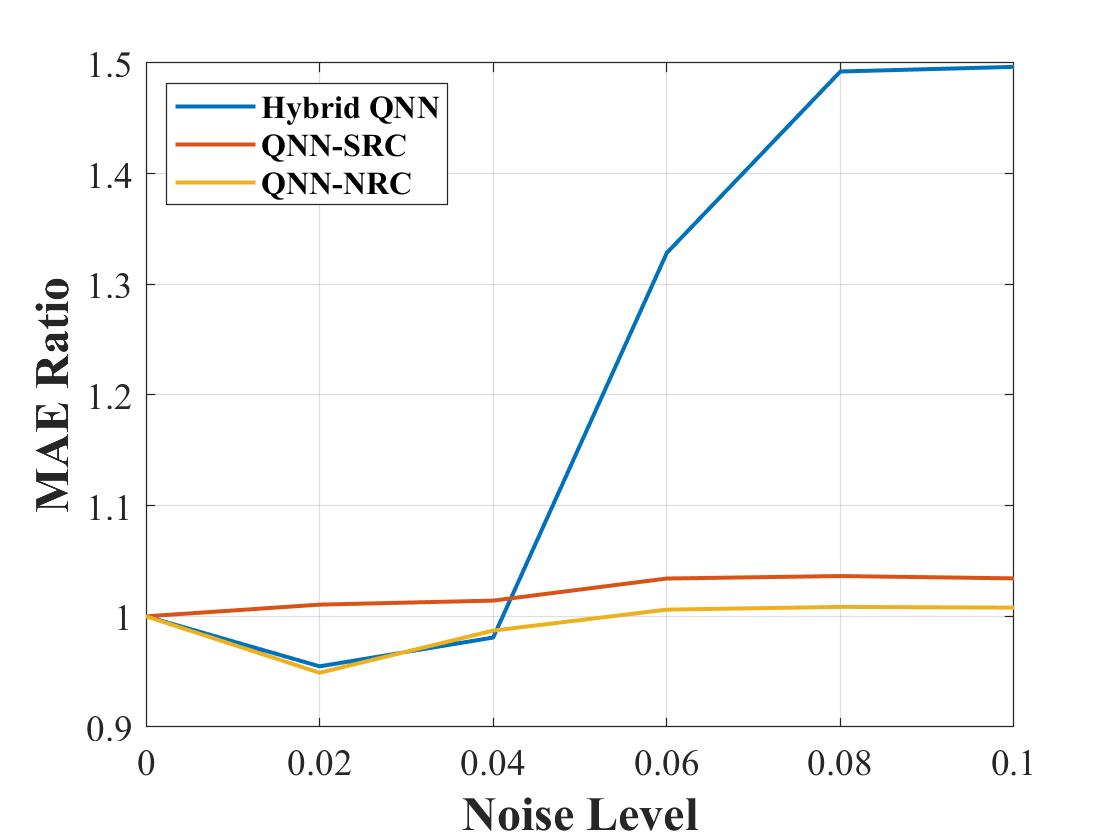} 
	\caption{Illustration of the impact of various noise levels on the performance of different QNNs on IEEE 39-bus system}
	\vspace{-4mm}
	\label{fig: noise}
\end{figure}

This subsection evaluates the sensitivity of the conventional Hybrid QNN and the proposed QNNs (QNN-SRC and QNN-NRC) to quantum noise using the IEEE 39-bus system. The results are presented in Fig.~\ref{fig: noise}, where the x-axis represents the overall quantum noise level $e$, ranging from $2\%$ to $10\%$, and the y-axis (MAE Ratio) denotes the normalized MAE, computed as the total MAE under a given noise level divided by the noise-free baseline MAE.
As shown in Fig.~\ref{fig: noise}, at low noise levels ($2\%-4\%$), Hybrid QNN and QNN-NRC exhibit a slight reduction in MAE, while QNN-SRC remains relatively stable. This behavior may be attributed to the phenomenon of noise-induced regularization, wherein moderate levels of quantum noise can suppress overfitting and enhance generalization in QNNs. However, beyond a noise level of 0.04, all methods exhibit an increasing trend in MAE, indicating a degradation in predictive accuracy as quantum noise intensifies.

The impact of noise is particularly pronounced for Hybrid QNN, where the MAE ratio escalates sharply, reaching approximately 1.5 at the highest noise level ($10\%$). This suggests that the Hybrid QNN is highly susceptible to quantum noise, likely due to cumulative stochastic and coherent errors affecting quantum circuit fidelity. In contrast, the proposed physics-informed residual-connected QNNs (QNN-SRC and QNN-NRC) exhibit substantially greater resilience, with MAE ratios remaining below 1.05 across all noise levels. This robustness can be attributed to residual connections that enhance gradient propagation and physics-informed constraints that mitigate error accumulation, thereby stabilizing optimization and inference in the presence of noise.
The results demonstrate that while all QNN architectures experience performance degradation with increasing quantum noise, the proposed QNNs (QNN-SRC and QNN-NRC) exhibit significantly enhanced robustness compared to the conventional Hybrid QNN. 

\section{Conclusion}
This paper presents a hybrid classical-quantum deep learning model for solving AC-OPF problems using a small number of qubits. To address the challenges associated with quantum circuit training, particularly the barren plateau problem, we propose two effective solutions for AC-OPF problems, validate the impact of classical neural network components in quantum machine learning, and emphasize the importance of residual learning and physics-informed design in QNN-based optimization tasks. 
The proposed methods are evaluated on multiple benchmark test systems, demonstrating significantly lower errors compared to conventional NNs and Hybrid QNN. The results further confirm the generalizability of the proposed models across various network topologies. Moreover, sensitivity analysis highlights the robustness of the proposed approaches under quantum noise, which is an essential property for implementation on NISQ devices.

\bibliographystyle{IEEEtran}
\bibliography{IEEEabrv,PWRSBib}

\end{document}